\documentstyle[aps,prl,epsfig]{revtex}

\begin{document}
\title{Autofeedback scheme for preservation of macroscopic coherence \\
in microwave cavities}
\author{M. Fortunato~$^*$, J.M. Raimond~$^\dagger$, 
P. Tombesi~$^*$ and D. Vitali~$^*$}
\address {$^*$Dipartimento di Matematica e Fisica, Universit\`a di
Camerino, via Madonna delle Carceri I-62032 Camerino \\
and INFM, Unit\`a di Camerino, Italy\\
$^\dagger$Laboratoire Kastler Brossel, D\'epartement de Physique
de l'Ecole Normale Sup\'erieure, \\ 
24 rue Lhomond, F-75231 Paris Cedex 05, France} 
\date{\today}
\maketitle

\begin{abstract}
We present a scheme for controlling the decoherence of 
a linear superposition of
two coherent states with opposite phases in a high-Q microwave cavity,
based on the injection of appropriately prepared ``probe'' and
``feedback'' Rydberg atoms,
improving the one presented in
[D. Vitali {\it et al.}, Phys. Rev. Lett. {\bf 79}, 2442 (1997)]. 
In the present scheme, the information transmission from the
probe to the feedback atom is directly mediated by a second auxiliary cavity. 
The detection efficiency for the probe atom is no longer a critical parameter, and
the decoherence time of the 
superposition state can be significantly increased using 
presently available technology.
\end{abstract}


\section{Introduction}

The problem of how the classical macroscopic world emerges from the quantum 
substrate is
an important point in the interpretation of quantum mechanics 
and it is still the subject of an intense debate \cite{zur,ZEH98}.  
This problem is well 
illustrated by the possibility, 
opened by 
quantum mechanics, of having linear 
superpositions of macroscopically distinguishable states, the 
so-called ``Schr\"odinger cat'' states. 
An explanation of 
why we never observe these paradoxical states is 
proposed by the {\em 
decoherence} models, i.e., the rapid transformation of these linear 
superpositions into the corresponding classical statistical mixture, 
caused by the unavoidable entanglement of the system
with uncontrolled 
degrees of freedom of the environment \cite{zur}. The decoherence time
depends on the form of system-environment interaction \cite{anglin} 
but, in most cases, it is inversely proportional to the squared 
``distance'' between the two states of the superposition 
\cite{milwal}.
For macroscopically distinguishable states, the 
decoherence process becomes 
thus practically instantaneous \cite{zur}. 
Decoherence is experimentally accessible only in the {\em 
mesoscopic} domain. In this 
case, one is able to monitor the progressive emergence of classical 
properties from the quantum ones. A first
important achievement has been obtained by Monroe {\it et
al.} \cite{wine}, who
prepared a trapped ${\rm ^{9}Be^{+}}$ ion
in a superposition of spatially separated coherent states and
detected the quantum coherence between the two localized states.
However, the decoherence of the 
superposition state has not been studied in this experiment. 
The progressive decoherence of a mesoscopic Schr\"odinger cat has been 
observed for the first time in the experiment of Brune  
{\it et al}. \cite{prlha}, where the linear
superposition of two coherent states of the electromagnetic field
in a cavity with classically distinct phases has been generated and 
detected.

With the impressive development of quantum information 
theory in the last years \cite{bennet}, 
the study of decoherence has become important not only from a 
fundamental, but also from a more practical point of view. 
All the quantum information processing applications
rely on the possibility of performing unitary 
transformations on a system of $N$ quantum bits, 
whose decoherence has to be made as small as possible. For this 
reason, decoherence control is now a rapidly expanding 
field of investigation. In this respect, quantum error correction codes 
\cite{error} have been developed in which
the entangled superposition state of $N$ qubits is 
``encoded'' in a larger number 
of qubits. Assuming that only a fraction of qubits decoheres, it 
is then possible to reconstruct the original state with a suitable
decoding procedure,
provided that errors affect different qubits independently. 
These codes always require the entanglement of a large number 
of qubits, and will become practical only if quantum networks of 
tens of qubits become available. Up to now, the polarization states
of three photons have been entangled at most \cite{ghz}.
Entangled states of two Rydberg atoms \cite{EPRPAIR} or
of two trapped ions \cite{turch} at most can be generated.
Therefore, in the 
present experimental situation, it is 
more realistic to study complementary 
and more ``physical'' ways to harness decoherence, based on the 
knowledge of the specific process causing decoherence, which could 
be applied with very few degrees of freedom. This is 
possible, in particular, in quantum optics, when information is 
encoded in the quantum states of an electromagnetic mode (see for
example \cite{milb}). In this case 
decoherence is caused by photon leakage.
It could therefore  be 
possible to develop experimental schemes able to face photon leakage 
and the associated decoherence.

A series of papers \cite{homo,minsk,prlno,jmo,pran} have shown that 
a possible way to control decoherence in optical cavities
is given by appropriately designed feedback schemes. Refs.~\cite{homo} 
show that a 
feedback scheme based on the continuous homodyne measurement of an 
optical cavity mode is able to increase the decoherence time of a 
Schr\"odinger cat state. In Ref.~\cite{jmo,pran} a feedback scheme 
based on continuous photodetection and the injection of appropriately 
prepared atoms has been considered. This scheme, in the limit
of very good detection efficiency, is able to obtain 
a significant ``protection'' of a generic quantum state in a cavity.
In \cite{prlno,pran} this photodetection-mediated scheme has been 
adapted to the microwave experiment of Ref.~\cite{prlha} in which 
photodetectors cannot be used.
The cavity state can only be indirectly 
inferred from measurements performed on probe atoms which have 
interacted with the cavity mode.
Under ideal conditions, this adaptation to the microwave cavity case 
leads to a 
significant increase of the lifetime of the Schr\"odinger cat generated 
in \cite{prlha}.
It suffers however from two 
important limitations, making it very inefficient when 
applied under the actual experimental situation. 
It first requires the preparation of samples containing {\em exactly} one
Rydberg atom sent through the apparatus. Up to now, the experimental 
techniques
allow only to prepare a sample 
containing a random atom number, with a Poisson statistics. 
Two-atom
events are excluded only at the expense of a low average atom number, 
lengthening
the feedback loop cycletime
\cite{EPRPAIR}. The original scheme requires also a 
near unity atomic detection efficiency,
which is extremely difficult to achieve even with the foreseeable 
improvements of the experimental apparatus.

In this paper we present a significant improvement 
of the microwave feedback scheme described in \cite{prlno,pran}. 
This new version, using
a direct transmission of the quantum information from the probe to the feedback 
atom,
does not require a large detection efficiency, removing one of the main 
difficulties of the previous 
design. It however also requires sub-poissonian atom statistics. 
We show briefly how such
atomic packets could be 
in principle prepared with standard laser techniques. 
Finally, our scheme improves the efficiency of the 
feedback photon injection in the cavity by using an adiabatic rapid passage.

The paper is organized as follows: in section II the feedback 
scheme of \cite{prlno,pran} is reviewed and critically discussed; in 
section III the modifications of this scheme are introduced and in 
section IV the map describing the feedback cycle is derived. 
In section V the dynamics in the presence of feedback is studied 
and the protection capabilities of the new proposal are illustrated, 
while section VI is devoted to concluding remarks.
 
\section{The feedback scheme based on atomic detection}

Let us briefly review the original ``stroboscopic'' feedback scheme
for microwave cavities proposed in \cite{prlno,pran}. 
This proposal is based on a very simple idea: 
whenever the cavity looses a 
photon, a feedback loop supplies the cavity mode with another 
photon, through the injection of an appropriately prepared atom.
However, since there are no good enough photodetectors for microwaves, 
one has to find an indirect way to check if the high-Q microwave 
cavity has lost a photon or not. In the experiment of Brune {\it
et al.} \cite{prlha}, information on the cavity field state is 
obtained by detecting the state of a 
circular Rydberg atom which has
dispersively interacted with the 
superconducting microwave cavity. This provides 
an  ``instantaneous" measurement of the cavity field
and suggests that
continuous photodetection can be replaced 
by a series of {\it repeated} measurements, performed by 
non--resonant atoms regularly crossing the high-Q  
cavity, separated by a time interval $\tau_{pr}$. 

The experimental scheme of the stroboscopic feedback loop is a
simple modification of the scheme employed in
Ref.~\cite{prlha}.
The relevant levels of the velocity-selected atoms
are two adjacent circular Rydberg states with principal 
quantum numbers $n=50$ and $n=51$ (denoted by 
$|g\rangle$ and $|e\rangle$ respectively) and a very long lifetime 
($30$ ms). 
The high-Q superconducting cavity is sandwiched between two low-Q 
cavities $R_{1}$ and $R_{2}$, 
in which classical microwave fields resonant 
with the transition between $|e\rangle$ and $|g\rangle$ can be applied. 

The high-Q cavity $C$ is instead slightly off-resonance 
with respect to the 
$e\, \rightarrow \, g$ transition, with a detuning
\begin{equation}
\delta = \omega - \omega_{eg}\;,
\label{detu}
\end{equation}
where $\omega $ is the cavity mode frequency and 
$\omega_{eg}=(E_{e}-E_{g})/\hbar $. The Hamiltonian of the 
atom-microwave cavity mode system is the usual Jaynes-Cummings Hamiltonian,
given by
\begin{eqnarray}
&&H_{JC}=E_{e}|e\rangle \langle e | + E_{g}|g\rangle \langle g |+
\hbar \omega a^{\dagger} a \nonumber \\
&&+\hbar \Omega \left(|e \rangle \langle g |a+|g \rangle \langle e |
a^{\dagger}\right) \;,
\label{jc}
\end{eqnarray}
where $\Omega $ is the vacuum Rabi coupling between the atomic dipole 
on the $e\, \rightarrow \, g$ transition and the cavity mode.
In the off-resonant case and perturbative limit $\Omega \ll \delta $, the 
Hamiltonian (\ref{jc}) (under an appropriate redefinition of 
level phases)
assumes the dispersive form \cite{pran,harray,brune}
\begin{equation}
H_{disp}=\hbar \frac{\Omega ^{2}}{\delta} \left(|g\rangle
\langle g| a^{\dagger} a-|e \rangle \langle e | 
a^{\dagger} a \right)\;.
\label{heff}
\end{equation}

The Schr\"odinger cat state is generated when the cavity mode is 
initially in a coherent state $|\alpha \rangle $ and the Rydberg atom, which is 
initially prepared in the state $|e\rangle $, is subjected to a $\pi/2$ 
pulse both in $R_{1}$ and in $R_{2}$. In fact, 
when the atom has left the cavity 
$R_{2}$, the joint state of the atom-cavity system becomes
the entangled state \cite{prlha,pran,brune} 
\begin{equation}
|\psi_{atom+field}\rangle = \frac{1}{\sqrt{2}}\left(|e\rangle 
\left(|\alpha e^{i\phi}\rangle -|\alpha e^{-i\phi}\right)
+|g\rangle \left(|\alpha e^{i\phi}\rangle +|\alpha e^{-i\phi}\right) \right)\;,
\label{atf1}
\end{equation}
where $\phi = \Omega ^{2}t_{int}/\delta $ and $t_{int}$ is the
interaction time in $C$.
A cat state, i.e. a linear superposition of two coherent states with
different phases, is then conditionally generated in the microwave cavity
as soon as one of the two circular atomic states is detected. 

As it was shown in Ref.~\cite{pran}, the stroboscopic feedback scheme
works only for Schr\"odinger cat states with a definite parity, i.e.
even or odd cat states, and therefore we shall restrict to $\phi = \pi/2$
from now on. In fact, when the cavity field initial state is a generic
density matrix $\rho$, the state of the probe atom-field system after
the two $\pi/2$ pulses and the $\phi=\pi/2$ conditional phase-shift can be
written as \cite{pran}
\begin{equation}
\rho_{atom+field}= |e\rangle \langle e| \otimes \rho_{e} +
|g\rangle \langle g| \otimes \rho_{g} 
+|e\rangle \langle g| \otimes \rho_{+} +
|g\rangle \langle e| \otimes \rho_{-} \;,
\label{2at3}
\end{equation}
where 
\begin{eqnarray}
\rho_{e} &=& P_{odd}\rho P_{odd}
\label{odd} \\
\rho_{g} &=& P_{even}\rho P_{even} \;,
\label{even} 
\end{eqnarray}
are the projections of the cavity field state onto the subspace 
with an odd and even number of photons, respectively, and the operators
$\rho_{\pm}$ (whose expression is not relevant here) 
are given in \cite{pran}.
 Eq.~(\ref{2at3}) shows that there is a perfect
correlation between the atomic state and the cavity field parity,
which is the first step in an optimal 
quantum non demolition measurement of the photon number \cite{OPTQND}.
It is possible to prove 
that this perfect correlation between the atomic
state and a cavity mode property holds only in the case of an exact 
$\phi=\pi/2$-phase shift 
sandwiched by two classical $\pi/2$ pulses in cavities
$R_{1}$ and $R_{2}$ \cite{pran}.
Moreover, the entangled state of Eq.~(\ref{2at3}) allows to 
understand how it is possible to check if the microwave cavity $C$ 
has lost a photon or not and therefore to trigger the feedback loop,
using atomic state detection only.
The detection of $e$ or $g$ determines the parity of
the field and, provided that the probe atomic pulses are
frequent enough,
indicates whether a microwave photon 
has left $C$ or not.
In fact, let us consider for example the case in which an odd cat 
state is generated (first atom detected in $e$):  
a probe atom detected in state
$e$ means that the cavity field has remained in the odd 
subspace.
The cavity has therefore lost an {\it even} 
number of photons.  If the time interval $\tau_{pr}$ between
the two atomic pulses is much smaller than the 
cavity decay time $\gamma^{-1}$, $\gamma \tau_{pr} \ll 1$,
the 
probability of loosing two or more photons is negligible and 
this detection of the probe atom in $e$ means that no photon has leaked out 
from the high-Q cavity $C$.  On the contrary, when the probe atom is 
detected in $g$, the cavity mode state is projected into the even 
subspace.
The cavity has then lost an 
{\it odd} number of photons.  Again, in the limit of enough closely 
spaced sequence of probe atoms, $\gamma \tau_{pr} \ll 1$, the probability of 
loosing three or more photons is negligible.
A detection in $g$ means that one photon has exited the cavity.
Therefore, for achieving a good protection of the initial odd cat 
state, the feedback loop
has to supply the superconducting cavity with a photon whenever the 
probe atom is detected in $g$, while feedback must not act 
when the atom is detected in the $e$ state.

In Ref.~\cite{pran} it has been proposed to realize this
feedback loop with a switch connecting the 
$g$ state field-ionization detector with a second atomic injector, sending 
an atom in the excited state $e$ into the high-Q cavity. 
The feedback atom is put in resonance with the cavity mode by another 
switch turning on an electric field in the cavity $C$ when the atom enters it, 
so that the level $e$ is Stark-shifted into resonance with the cavity mode.

As it is shown in Ref.~\cite{pran}, if the probe atomic pulses
are sufficiently frequent, this stroboscopic feedback scheme becomes
extremely efficient and one gets a good preservation of an initial 
Schr\"odinger cat state. However, if we consider the adaptation of this 
scheme to the present experimental apparatus of Ref.~\cite{prlha}, we 
see that it suffers from two main limitations, which significantly 
decrease its efficiency. First of all the scheme is limited 
by the non-unit efficiency of the atomic state detectors 
($\eta_{det}\simeq 0.4$), since the 
feedback loop is triggered only when the $g$-detector clicks. Most 
importantly, the above scheme assumes 
one has perfect ``atomic 
guns'', i.e. the possibility of having probe and feedback atomic 
pulses with {\it exactly one atom}.
This is not experimentally achieved up to now. The actual experiment \cite{prlha} has 
been performed using atomic pulses with a probability of having 
exactly one atom $p_{1} \simeq 0.2$, 
close to the mean atom number in the sample.
This low mean atom number has been chosen to minimize two--atom events. In this 
experimental situation, the proposed stroboscopic feedback scheme would 
have an effective efficiency 
$\eta_{eff}=\eta_{det}p_{1}^{2} \simeq 0.016$, 
too low to 
get an appreciable protection of the Schr\"odinger cat state. 
We show here how this scheme may be improved and adapted to the experimental apparatus employed 
in Ref.~\cite{prlha}.

\section{The new stroboscopic feedback loop}

The limitations due to the non-unit efficiency of the atomic detectors 
could be avoided if we eliminate the measurement step in the feedback 
loop and replace it with 
an ``automatized'' 
mechanism
preparing the correct feedback atom whenever 
needed. This mechanism can be provided by an appropriate conditional 
quantum dynamics. 
We need a ``controlled-NOT" gate between 
the probe atom and the feedback atom, because the feedback atom has 
to remain in an ``off''state if the probe atom exits the cavity of 
interest $C$ in the $e$ state, while the feedback atom has to be in 
the excited state $e$ when the probe atom leaves $C$ in the $g$ state
(we are still assuming 
the initial generation of an odd cat state). 
This conditional dynamics can be provided by a second 
high-Q microwave cavity $C'$, similar to $C$, replacing the atomic 
detectors,
crossed by the probe atom first and by the 
feedback atom soon later.
A schematic description of the new feedback scheme, with the second 
cavity $C'$ replacing the atomic state detectors is given by Fig.~\ref{appa}.

The cavity $C'$ is 
resonant with the transition between an auxiliary circular state $i$, 
which can be taken as the immediately 
lower circular Rydberg state $n=49$, and level $g$.
The 
interaction times have to be set so that both the probe and the 
feedback atom 
experience a $\pi$ pulse when they cross the empty cavity $C'$ in 
state $g$
(or when they enter in state $i$ with one photon in $C'$). 
This interaction copies the state
of the probe atom onto the cavity mode and back onto the feedback atom. 
$C'$ acts thus
as a ``quantum memory'' \cite{MEMORY}, transferring directly 
the quantum information 
between the two atoms without need of a detection. 
This removes thus any need for a unit detection efficiency.

This fine 
tuning of the interaction times 
to achieve the $\pi$--spontaneous emission pulse condition
 can be obtained applying  
through the superconducting mirrors of $C'$ appropriately shaped 
Stark-shift electric fields which 
puts the atoms in resonance with 
the cavity mode in $C'$ only for the desired time. In this 
way, since $C'$ is initially in the vacuum state, one has
\begin{eqnarray}
|e\rangle _{p} |0\rangle _{C'} &\rightarrow & |e\rangle _{p} |0\rangle _{C'}
\label{prob1} \\
|g\rangle _{p} |0\rangle _{C'} &\rightarrow & |i\rangle _{p} |1\rangle _{C'}
\label{prob2}
\end{eqnarray}
when the probe atom crosses $C'$; soon later a feedback atom enters 
$C'$ in the state $|i\rangle_{f}$ and one has
\begin{eqnarray}
|i\rangle _{f} |0\rangle _{C'} &\rightarrow & |i\rangle _{f} |0\rangle _{C'}
\label{fb1} \\
|i\rangle _{f} |1\rangle _{C'} &\rightarrow & |g\rangle _{f} |0\rangle _{C'}
\label{fb2} 
\end{eqnarray}
(the cavity has a very high Q and therefore the probability of 
photon leakage in the meanwhile
is negligible). In this way the cavity $C'$ is always 
left disentangled in the vacuum state. 
The feedback atom exiting $C'$ in $|g\rangle $ can be promoted to 
$|e\rangle $ before 
entering $C$, as required by the feedback scheme, by subjecting it to
a $\pi$ pulse in the classical cavity $R_{2}$ (see Fig.~1).
The conditional dynamics provided by $C'$ eliminates any limitation
associated to the measurement and leads to an ``automatic feedback'' 
scheme with 
unit efficiency in principle.

As mentioned above, an important limitation of the stroboscopic feedback scheme
of \cite{prlno,pran} is that it requires exactly one probe 
and one feedback atom 
per loop. 
This condition is still
needed in the new scheme with the cavity $C'$ 
replacing the atomic detectors. With two 
or more probe atoms simultaneously in $C$ and in $C'$, one
gets a wrong phase shift for the field in 
$C$ and also an incomplete excitation transfer from the probe atom to 
the field in $C'$. The same condition holds for the feedback atoms. 
With two or more feedback atoms in the 
sample, the excitation transfers in $C'$ and $C$ are 
incomplete. 

A better control of the atom number, providing single atom events 
with a high probability, could be achieved
by a modification of the Rydberg atoms preparation techniques. 
We outline here briefly
the method, which could be implemented in a future version of the 
experimental set--up. The Rydberg atoms preparation would 
start from a very low--intensity velocity--selected 
Rubidium atomic beam. 
The ground state atom density is so low that
the average distance between the atoms in the beam is of the order 
of a few millimeters. It means that a section of the beam
a few millimeters long contains on the average only one atom 
(with a Poisson statistics). This section could be driven 
by a laser resonant on the 5S to 5P transition
(see Fig.~\ref{levels} for a schematic diagram of the relevant $^{85}$Rb
energy levels involved).
The fluorescence signal should make it possible to distinguish easily
the situations where the probed section of the beam contains zero, 
one, two or more atoms, implementing an {\em atom counter}. 
When the section contains 
zero, two or more atoms, it is discarded. The system waits then for a time 
$\tau_{pr}$
(a few microseconds to twenty microseconds, depending upon 
the atomic velocity and the precise length of the atomic beam section) 
until a fresh section of the beam comes in the probe laser beam. At variance, 
if the fluorescence level corresponds to exactly one
atom, the circular state preparation is started. 
Using only adiabatic rapid passages, it should 
be possible to promote the single ground state atom to the 
desired circular state with a high probability. 
The circular state preparation \cite{CIRCULAR,jmobru} 
proceeds in two steps. First, 
a laser excitation of an ``ordinary" Rydberg state,
then a transfer to the circular state. The latter step
already uses adiabatic rapid passages and has a very high efficiency. 
The former step could also be adiabatic, by using higher 
laser powers readily available.

Instead of preparing a random atom number at a given time, one thus
prepares with a high probability a single Rydberg atom after a random delay 
(since the preparation step is triggered only when the atomic counter gives 
a count of exactly one). The average value of this random delay is minimal 
when the probability to have exactly one atom is maximized.
With a Poissonian statistics, the
optimal mean number of atoms in the probed section is 1. 
The average random delay could be of the order of
25 $\mu$s in realistic experimental conditions. This is short enough 
at the scale of the cavity field lifetime 
to play no major role in the experiment. The unavoidable 
imperfections of the circular 
state preparation could be easily taken into account by assuming that
the sample contains one atom with a probability $p_r$ and 
no atoms with a probability 
$1-p_r$. Two-atom events are 
excluded, a considerable improvement 
compared to other preparation methods.

The timing of the whole experiment should be conditioned to the 
operation of the atomic counters. When the cycle starts,
the system idles until a probe atom has been counted and 
prepared in the circular state. After it has crossed $C'$,
the preparation cycle of the feedback atom is started. The system also 
idles until this atom is counted and prepared in the
circular state. The feedback is complete when this feedback atom has 
crossed the cavity $C$.

\section{The feedback cycle in more detail}   

Let us now determine the map of a generic feedback cycle, that is, the 
transformation connecting the states $\rho_{m}$ 
and $\rho_{m+1}$ of the cavity field in $C$ soon 
after the passage of two successive feedback atoms in $C$.
From the previous section 
it is clear that a new cycle 
begins only when one is sure to have one probe atom with certainty 
and therefore one has to wait a random time $t_{r}$ before the new 
probe atom enters $C$. 

The atomic counter 
operate with a cycle time $\tau_{pr}$. The average
number of atoms per probed packet being one, the 
probability of having exactly one atom 
is $p_{1}=1/e=0.37$. Therefore, the random waiting 
time can be written as
$t_{r}=l \tau_{pr}$, $l=0,1,\ldots$, where the probability 
distribution of the discrete random variable $l$ is given by
\begin{equation}
	p(l)=p_{1}(1-p_{1})^{l}\;\;\;\;\;\;l=0,1,\ldots \;.
	\label{prol}
\end{equation}
The first step of the feedback loop is simply the standard 
dissipative evolution with damping rate $\gamma$ 
for a random time $l\tau_{pr}$ 
\cite{pran,herzog}
\begin{equation}
\rho_{m}^{I}=\sum_{k=0}^{\infty}A_{k}(l\tau_{pr})\rho_{m} 
A_{k}(l\tau_{pr})^{\dagger}\;,
\label{step1}
\end{equation}
where 
\begin{equation}
A_{k}(t)=\sum_{n=0}^{\infty}\sqrt{\frac{(n+k)!}{n! k!} e^{-n \gamma 
t}\left(1-e^{-\gamma t}\right)^{k}} |n\rangle \langle n+k |  
\label{cnk}
\end{equation}
and where $\rho_{m}^{i}$ will denote the state after the 
$i$-th step of the cycle. 

\begin{figure}
\centerline{\epsfig{figure=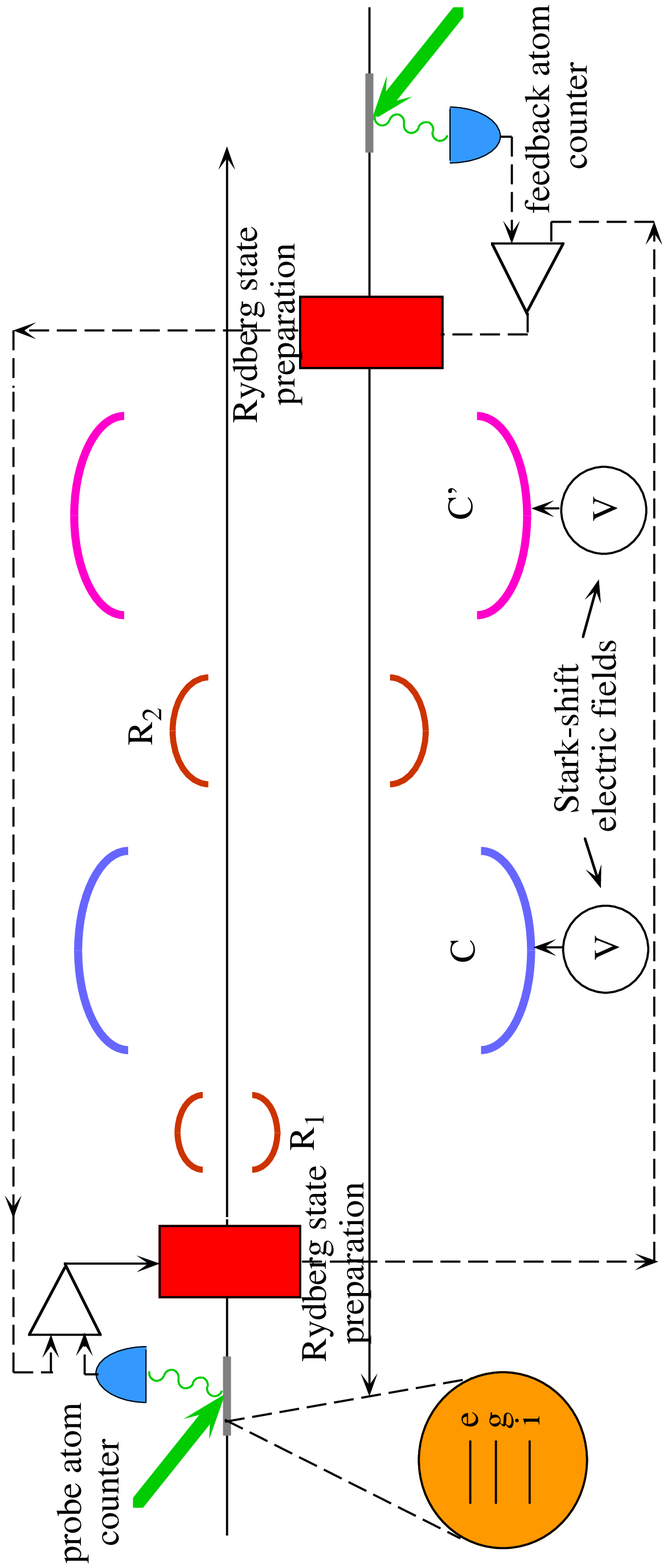,width=12cm,angle=270}}
\caption{Schematic diagram of the autofeedback scheme
proposed in this paper. $R_{1}$ and $R_{2}$ are the two 
cavities in which classical microwave pulses can be applied, $C$ is 
the microwave cavity of interest and $C'$ is the cavity automatically 
performing the needed correction. Electric fields can be applied at 
the superconducting mirrors of $C$ and $C'$ to Stark shift the 
Rydberg levels in order to tune the interaction times in $C'$ and 
realize adiabatic transfer in $C$.}
\label{appa}
\end{figure}

\begin{figure}
\centerline{\epsfig{figure=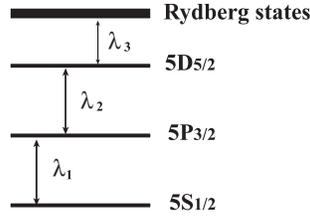,width=4cm}}
\caption{Schematic diagram of the relevant $^{85}$Rb energy levels involved
         in the atom counting and in the preparation of the circular 
         Rydberg states. The transition between the 5S$_{1/2}$ ($F=2, 
         3$) ground state  and the 5P$_{3/2}$ ($F=1, 2, 3, 4$) first
         excited state is driven by a laser diode at $\lambda_{1}= 780$ nm
         and is used for the fluorescence detection. The corresponding
         cycling transition is between ($F=3, m_{\rm F}=3$) and
         ($F'=4, m_{\rm F}'=4$). The transition between the first 
         and the 5D$_{5/2}$ second excited state is driven by a laser
         diode at $\lambda_{2}=776$ nm. Finally, the transition 
         between the second excited state and the Rydberg states is 
         driven by a laser diode at $\lambda_{3}=1.26$ $\mu$m.}
\label{levels}
\end{figure}

The second step is determined by the probe atom crossing the cavity 
$C$ and interacting with it via the dispersive Hamiltonian 
(\ref{heff}). Since the probe atom has been already prepared in the 
circular state $e$ and it has already crossed the classical 
cavity $R_{1}$ (see Fig.~\ref{appa}), 
it enters $C$ in the state $(|e\rangle +|g\rangle 
)/\sqrt{2}$. Due to the $\pi/2$ phase shift, 
the cavity mode in $C$ gets entangled with the probe atom and, after 
the second feedback step, one has
\begin{eqnarray}
&&\rho_{m}^{II}= \frac{1}{2}\left(
|e\rangle \langle e| \otimes P\rho_{m}^{I}P +
|g\rangle \langle g| \otimes \rho_{m}^{I} \right. \nonumber \\
&& \left. +|e\rangle \langle g| \otimes P\rho_{m}^{I} +
|g\rangle \langle e| \otimes \rho_{m}^{I}P \right)\;,
\label{step2}
\end{eqnarray}
where $P=\exp\left\{i\pi a^{\dagger}a\right\}$ is the cavity 
mode parity operator.  

As it can be seen from Fig.~\ref{appa}, the probe atom flies then 
from the cavity $C$ to the second high-Q cavity $C'$. During this time 
of flight one has to consider the effect of standard vacuum damping on 
the $C$ cavity mode and also the effect of the $\pi/2$ pulse in 
$R_{2}$ on the probe atom, yielding
\begin{eqnarray}
|e\rangle & \rightarrow & \frac{1}{\sqrt{2}}\left(|e\rangle 
+|g\rangle\right) \nonumber \\
|g\rangle & \rightarrow & \frac{1}{\sqrt{2}}\left(-|e\rangle 
+|g\rangle\right) \;.
\label{pi2}
\end{eqnarray}    
Note that we shall always neglect the spontaneous decay of the circular 
levels, since the lifetime of the involved level (about $30 $
ms) is much larger than the mean feedback cycle duration time
(of the order of $1$ ms).
The two actions do not interfere and therefore, rearranging the terms, 
one has an expression connected to Eq.~(\ref{2at3}) 
\begin{eqnarray}
&&\rho_{m}^{III}= \sum_{k=0}^{\infty}A_{k}(t_{C\rightarrow C'})
\left(
|e\rangle \langle e| \otimes \rho_{e}^{I} +
|g\rangle \langle g| \otimes \rho_{g}^{I} \right. \nonumber \\
&& \left. +|e\rangle \langle g| \otimes \rho_{+}^{I} +
|g\rangle \langle e| \otimes \rho_{-}^{I}\right) 
A_{k}(t_{C\rightarrow C'})^{\dagger} \;,
\label{step3}
\end{eqnarray}
where the density matrices $\rho_{e}^{I}$ and $\rho_{g}^{I}$ are 
the odd and even projections of Eqs.~(\ref{odd}) and (\ref{even}) and 
\begin{equation}
\rho_{\pm}=\frac{1}{4}\left[P\rho P - \rho \pm P \rho \mp \rho 
P\right] \;.
\end{equation}

The fourth step is determined by the interaction of the probe atom with 
the second high-Q cavity $C'$, 
which is described by the resonant interaction between the $C'$ 
cavity mode and the two lower circular levels $i$ and $g$
\begin{equation}
H_{C'}=\hbar \Omega' \left(|g \rangle \langle i |b +|i \rangle \langle 
g |b^{\dagger}\right) \;,
\label{cprime}
\end{equation}
where $\Omega'$ is the corresponding vacuum Rabi frequency and $b$ 
denotes the annihilation operator of the $C'$ cavity mode. The 
cavity $C'$ is initially in the vacuum state, and therefore, using the 
Stark tuning mechanism described in the preceding section  
to determine the effective interaction time
$t_{pr}^{int}$, it is 
possible to impose the $\pi$ pulse condition for $t_{pr}^{int}$ 
\begin{equation}
\Omega' t_{pr}^{int} = \frac{\pi}{2} \;,
\label{pipul}
\end{equation}
so that the conditional dynamics described by Eqs.~(\ref{prob1}) and 
(\ref{prob2}) is obtained. One gets therefore the following entangled state 
between the probe atom and the two microwave cavity modes
\begin{eqnarray}
&&\rho_{m}^{IV}= \sum_{k=0}^{\infty}A_{k}(t_{C\rightarrow C'})
\left(
|e\rangle _{p}\langle e| \otimes \rho_{e}^{I} \otimes |0\rangle _{C'}\langle 0|+
|g\rangle _{p}\langle g| \otimes \rho_{g}^{I}\otimes |1\rangle _{C'}\langle 1|
\right. \nonumber \\
&& \left. +|e\rangle _{p}\langle g| \otimes \rho_{+}^{I} 
\otimes |0\rangle _{C'}\langle 1|+
|g\rangle _{p}\langle e| \otimes \rho_{-}^{I} \otimes |1\rangle _{C'}\langle 0|
\right) A_{k}(t_{C\rightarrow C'})^{\dagger} \;.
\label{step4}
\end{eqnarray}
However the probe atom is not observed after exiting $C'$ and 
therefore we have to trace over it; as a result, the off-diagonal 
terms vanish and the following correlated state between the two 
microwave cavity modes is left 
\begin{equation}
\rho_{m}^{IV}= \sum_{k=0}^{\infty}A_{k}(t_{C\rightarrow C'})
\left(
\rho_{e}^{I} \otimes |0\rangle _{C'}\langle 0|+
\rho_{g}^{I}\otimes |1\rangle _{C'}\langle 1|
\right) A_{k}(t_{C\rightarrow C'})^{\dagger} \;.
\label{step4pr}
\end{equation}

During the probe atom crossing, 
the beam of feedback atoms continues to pass through
the apparatus in the opposite direction (see Fig.~\ref{appa}) in
their internal ground state, which is decoupled from all the microwave 
cavities of the experimental arrangement. Then the electronics 
controlling the circular state preparation of the feedback atom is set 
in such a way that one feedback atom can enter the cavity $C'$ in the 
Rydberg state $i$ soon after the probe atom has left it. However, as 
it happens for the probe atoms at the beginning of the cycle, one has 
to wait a random time until we are sure to have one feedback atom 
with certainty. 

We assume that also the feedback atoms are sent and counted with a time cycle 
$\tau_{fb}$. The probability of 
having one atom in a probed sample of the beam is again equal 
to $p_{1}=1/e=0.37$. 
Therefore, the random waiting time can be written as
$q \tau_{fb}$, $q=0,1,\ldots$, where $q$ is a 
discrete random variable with the same probability distribution
of the probe random variable $l$, given by Eq.~(\ref{prol}).
During this random waiting time, one has to consider
standard vacuum damping for both microwave cavities $C$ and $C'$. 
Photon leakage in $C'$ is particularly disturbing because it 
transforms the one photon state $|1\rangle \langle 1|$ into the 
vacuum, according to
\begin{equation}
|1\rangle \langle 1| \rightarrow e^{-\gamma' q \tau_{fb}}
|1\rangle \langle 1| + \left(1-e^{-\gamma' q \tau_{fb}}\right)
|0\rangle \langle 0| \;,
\label{vacua}
\end{equation}
($\gamma'$ is the cavity $C'$ damping rate)
blurring therefore any difference between the $C$ cavity 
states $\rho_{e}$ (that does not need any correction) and $\rho_{g}$
(that needs a photon back) in Eq.~(\ref{step4pr}). 
Using Eq.~(\ref{step4pr}), the resulting transformation for the joint 
state of the two microwave cavities becomes
\begin{eqnarray}
&&\rho_{m}^{V}= \sum_{k=0}^{\infty}A_{k}(t_{C\rightarrow C'}+q \tau_{fb})
\left\{\left[
\rho_{e}^{I} +\left(1-e^{-\gamma' q \tau_{fb}}\right) \rho_{g}^{I}\right]
\otimes |0\rangle \langle 0|  \right. \nonumber \\
&& \left. 
+e^{-\gamma' q \tau_{fb}} \rho_{g}^{I}\otimes |1\rangle \langle 1|
 \right\} A_{k}(t_{C\rightarrow C'}+q \tau_{fb})^{\dagger} \;.
\label{step5}
\end{eqnarray}

The next step of the feedback cycle is given by the resonant 
interaction of the feedback atom with the cavity mode $C'$. The 
interaction is again described by the Hamiltonian (\ref{cprime}) and 
one can use again the Stark-effect tuning mechanism to determine the 
right interaction time to get
the $\pi$ pulse condition of Eq.~(\ref{pipul}). The consequent 
transformation is described by Eqs.~(\ref{fb1}) and (\ref{fb2}), so 
that, after the feedback atom passage, the cavity $C'$ 
comes back to its initial vacuum state and the entanglement with the 
cavity of interest $C$ is transferred to the feedback atom.
Actually, in the 
preceding steps we have neglected the effect of photon leakage out of 
$C'$ during 
both probe and feedback atom passages through $C'$ 
because of its high Q value. We can 
partially amend this approximation by ``postponing'' dissipation 
after the interactions and adding the probe and feedback atom crossing times 
$t_{pr}^{cr}$ and $t_{fb}^{cr}$ 
to the random waiting time $q\tau_{fb}$ in (\ref{vacua}). The 
resulting $C$-field plus feedback atom joint state becomes
\begin{eqnarray}
&&\rho_{m}^{VI}= \sum_{k=0}^{\infty}A_{k}(t_{C\rightarrow C'}+q \tau_{fb})
\left\{\left[
\rho_{e}^{I} +\left(1-e^{-\gamma' (q 
\tau_{fb}+t_{pr}^{cr}+t_{fb}^{cr})}\right) \rho_{g}^{I} \right]
\otimes |i\rangle_{f} \langle i|  \right. \nonumber \\
&& \left. 
+e^{-\gamma' (q 
\tau_{fb}+t_{pr}^{cr}+t_{fb}^{cr})} \rho_{g}^{I}\otimes |g\rangle_{f} \langle 
g| \right\} A_{k}(t_{C\rightarrow C'}+q \tau_{fb})^{\dagger} \;.
\label{step6}
\end{eqnarray}

As we have explained in the preceding section, the odd density matrix
$\rho_{e}^{I}$ does not need any correction and therefore has to be 
correlated with $|i\rangle _{f}$, while the even part $\rho_{g}^{I}$ 
needs a correction and therefore has to be correlated with $|g\rangle _{f}$.
Looking at Eq.~(\ref{step6}), it is easy to see that the factor
$\exp\{-\gamma'\left(q \tau_{fb}+t_{pr}^{cr}+t_{fb}^{cr}\right)\}$ 
gives the probability that the feedback loop is acting correctly, 
i.e., this factor plays exactly the same role of the detector 
efficiency 
in the original stroboscopic feedback scheme of Refs.~\cite{prlno,pran}. 
However the times $\tau_{fb}$, $t_{pr}^{cr}$ and $t_{fb}^{cr}$ are 
very small in the experiment (of the order of $10$ $\mu$sec) and using 
a very high Q cavity for $C'$, i.e. 
$\gamma'\left(q \tau_{fb}+t_{pr}^{cr}+t_{fb}^{cr}\right) \ll 1$,
one obtains a feedback loop with an effective unit efficiency, 
which, as we have remarked in the preceding section, is one of the 
improvements of the new feedback scheme.

Then the feedback atom flies from $C'$ to $C$ and, along its path, it 
passes through the cavity $R_{2}$, within which it is subjected to a $\pi$ 
pulse on the transition $ g \rightarrow e$. The effect of this pulse 
is simply to transform the state $|g\rangle _{f}$ into $|e\rangle 
_{f}$ in Eq.~(\ref{step6}) and it does not interfere with the effect 
of vacuum damping on the $C$ cavity mode. 
Therefore it is easy to see that the state of Eq.~(\ref{step6}) is 
simply changed to
\begin{eqnarray}
&&\rho_{m}^{VII}= \sum_{k=0}^{\infty}A_{k}(t_{0}+q \tau_{fb})
\left\{\left[
\rho_{e}^{I} +\left(1-e^{-\gamma' (q 
\tau_{fb}+t_{pr}^{cr}+t_{fb}^{cr})}\right) \rho_{g}^{I} \right]
\otimes |i\rangle_{f} \langle i|  \right. \nonumber \\
&& \left. 
+e^{-\gamma' (q 
\tau_{fb}+t_{pr}^{cr}+t_{fb}^{cr})} \rho_{g}^{I}\otimes |e\rangle_{f} \langle 
e| \right\} A_{k}(t_{0}+q \tau_{fb})^{\dagger} \;,
\label{step7}
\end{eqnarray}
where $t_{0}$ is the overall time of flight, i.e. the sum of the probe 
atom time of flight from $C$ to $C'$ and the feedback atom time of 
flight from $C'$ to $C$.

We finally arrive at the last step of the feedback cycle, i.e. the 
interaction between the feedback atom and the cavity mode we want to 
protect against decoherence. If the feedback atom is in state 
$|i\rangle $ nothing relevant happens and the $C$ cavity mode state is 
left unchanged, as it must be. If instead the feedback atom is in 
state $|e\rangle $, it has to release its excitation to the cavity 
mode. In Ref.~\cite{prlno,pran} it has been proposed to realize
this excitation transfer by Stark-shifting into resonance the 
circular levels in order to use the resonant Jaynes-Cummings interaction
[Eq.~(\ref{jc}) with zero detuning $\delta$]. Here we propose 
to use the Stark tuning mechanism in a more clever way, in order to optimize
the photon transfer to the microwave cavity mode. In fact, if one 
uses the resonant interaction, the excitation transfer to the cavity 
is optimal for an odd number of half Rabi oscillations, that is
\begin{equation}
\Omega t_{fb}^{int} \sqrt{n+1}=\pi (m+1/2)\;\;\;\; m \;\;{\rm 
integer}\;.
\label{sqreso}
\end{equation}
The dependence of this condition on the intracavity
photon number $n$ is a limitation of the resonant interaction because 
the photon transfer becomes ideal in the case of a previously known 
Fock state only. On the contrary it would be preferable to have a way to 
perfectly release the photon in $C$ whatever the state of the cavity mode is. 
As explained in \cite{LEPAPE,adia} this possibility is provided by 
{\em adiabatic transfer}, which can be realized in the present 
context using a Stark shift electric field in $C$ able to change 
adiabatically the atomic frequency $\omega_{eg}$ through the resonant 
value $\omega$. 

Let us see in more detail how it is possible to use the Stark effect to 
realize the adiabatic transfer of the excitation. Let us consider the 
Hamiltonian of the Jaynes-Cummings model (\ref{jc}) in the interaction 
picture and with a time-dependent detuning $\delta(t) $ 
because of the adiabatic time dependence of the atomic frequency 
$\omega_{eg}$,
\begin{equation}
H_{ad}=\hbar \delta(t) a^{\dagger} a+
\hbar \Omega \left(|e \rangle \langle g |a +|g \rangle \langle 
e |a^{\dagger}\right) \;.
\label{hadia}
\end{equation} 
This Hamiltonian 
couples only states within the two-dimensional manifold 
with $n+1$ excitations spanned by 
$|g,n+1\rangle $ and $|e,n\rangle $, where $n$ denotes 
a Fock state of the cavity mode. Within this 
manifold one has the adiabatic eigenvalues 
\begin{equation}
\frac{E_{\pm}^{n}(t)}{\hbar}=\delta(t) \left(n+\frac{1}{2}\right)\pm 
\sqrt{\frac{\delta^{2}(t)}{4}+\Omega^{2}(n+1)}
\label{autova}
\end{equation}
and the corresponding adiabatic eigenstates
\begin{equation}  
|v_{\pm}^{n}(t)\rangle = N_{\pm}\left\{\left[
- \frac{\delta(t)}{2}\pm 
\sqrt{\frac{\delta^{2}(t)}{4}+\Omega^{2}(n+1)}\right]|e,n\rangle
+\Omega \sqrt{n+1}|g,n+1\rangle \right\} \;.
\label{vadia}
\end{equation}
Now, according to the adiabatic theorem \cite{messiah}, 
when the evolution from time $t_{0}$ to time $t_{1}$ is sufficiently slow, a 
system starting from an eigenstate of $H(t_{0})$ will pass into the 
corresponding eigenstate of $H(t_{1})$ that derives from it by continuity.
In the present case, the interesting adiabatic eigenstate is
$|v_{+}^{n}(t)\rangle $. In fact, if we assume that 
the detuning $\delta$ is varied
adiabatically from a large negative value $-\delta_{0}$
to a large positive value $\delta_{0}$, with 
$\delta_{0} \gg \Omega \sqrt{n+1}$, it is easy to see from 
Eq.~(\ref{vadia}) that 
$|v_{+}^{n}(t)\rangle $ will consequently show the following adiabatic 
transformation 
\begin{equation}
|e,n\rangle \rightarrow |g,n+1\rangle \;\;\;\;\;\;\; \forall n
\label{trasfe}
\end{equation}
thereby realizing the desired excitation transfer regardless of the
cavity mode state, which, in terms of cavity mode density matrices 
can be written as
\begin{equation}
\rho \rightarrow a^{\dagger} \frac{1}{\sqrt{a a^{\dagger}}} \rho
\frac{1}{\sqrt{a a^{\dagger}}} a \;.
\label{trasfe2}
\end{equation}
To be more precise, each adiabatic eigenstate gets
its own dynamical phase factor 
\begin{equation}
e^{-i\Phi_{n}}=e^{-\frac{i}{\hbar} \int dt E_{n}(t)}
\label{dyna}
\end{equation}
during the adiabatic evolution \cite{messiah} and therefore
the transformation (\ref{trasfe2}) 
exactly holds only if this dynamical phase factor does not
depend on $n$. Assuming a linear 
sweep of the Stark-shift electric field, that is,
$\delta(t) = \delta_{0} t/t_{s}$, for 
$|t| \leq t_{s}$ and using Eq.~(\ref{autova}), one has
\begin{equation}
\Phi_{n}=-\frac{\delta_{0} t_{s}}{2 \hbar}\left[\sqrt{1+
\frac{4 \Omega^{2} (n+1)}{\delta_{0}^{2}}}+
\frac{2\Omega^{2} (n+1)}{\delta_{0}^{2}} \log\left(
\frac{\sqrt{1+ 4 \Omega^{2} (n+1)/\delta_{0}^{2}}+1}
{\sqrt{1+ 4 \Omega^{2} (n+1)/\delta_{0}^{2}}-1}\right)\right] \;.
\label{fasen}
\end{equation}
Therefore one has in general a photon number dependent phase-shift; 
however in the particular adiabatic transformation we are considering,
for which $\delta_{0} \gg \Omega \sqrt{n+1}$ for all the relevant values
of $n$, this phase factor can be well approximated, at the lowest order
in $\Omega \sqrt{n+1}/\delta_{0}$, by the constant phase factor
$\exp\{i\delta_{0}t_{s}/2\hbar\}$ and therefore Eq.~(\ref{trasfe2})
holds exactly.

Finally we have all the ingredients to determine the last step of the
feedback cycle. One has to consider the transformation (\ref{trasfe2})
when the feedback atom is in state $e$, while nothing happens when the 
feedback atom passes $C$ in state $i$ and then one has to trace over the 
feedback atom because it is not observed. We have therefore the map
connecting the state of the $C$ cavity mode after two successive 
cycles, which is
\begin{eqnarray}
&&\rho_{m+1}= \sum_{k=0}^{\infty}A_{k}(t_{0}+q \tau_{fb})
\left[
\rho_{e}^{I} +\left(1-e^{-\gamma' (q 
\tau_{fb}+t_{pr}^{cr}+t_{fb}^{cr})}\right) \rho_{g}^{I} \right]
A_{k}(t_{0}+q \tau_{fb})^{\dagger} \nonumber \\
&&  +e^{-\gamma' (q 
\tau_{fb}+t_{pr}^{cr}+t_{fb}^{cr})} a^{\dagger} \frac{1}{\sqrt{a a^{\dagger}}}
\sum_{k=0}^{\infty}A_{k}(t_{0}+q \tau_{fb})\rho_{g}^{I}
 A_{k}(t_{0}+q \tau_{fb})^{\dagger} \frac{1}{\sqrt{a a^{\dagger}}} a\;,
\label{step8}
\end{eqnarray}
where the projected matrices $\rho_{e}^{I}$ and $\rho_{g}^{I}$
are obtained from the cavity mode state after the preceding
feedback cycle $\rho_{m}$ by inserting Eqs.~(\ref{odd}) and (\ref{even}) 
into Eq.~(\ref{step1}).

In the determination of the map (\ref{step8})
we have assumed that the Rydberg state
preparation for both probe and feedback atoms has unit efficiency.
In a realistic situation, the circular state preparation will have
a non-unit  efficiency $p_r <1$. This implies that the feedback map of 
Eq.~(\ref{step8}) is realized with a probability $p_{r}^{2}$ only. In fact,
when either the probe or feedback atom Rydberg state preparation fails,
the feedback does not effectively take place, because either the probe 
or the feedback atom is not in the correct state and the photon transfer 
in $C$ cannot take place.
This effect can be taken into account modifying the feedback map 
of Eq.~(\ref{step8}) in this way
\begin{equation}
\rho_{m+1}=p_{r}^{2} \Phi_{q,l}^{fb}\left(\rho_{m}\right)+
(1-p_{r}^{2}) \Phi_{q,l}^{diss}\left(\rho_{m}\right)
= \Phi_{q,l}\left(\rho_{m}\right) \;,
\label{stepind}
\end{equation}
where $\Phi_{q,l}^{fb}$ is the map operator defined in Eq.~(\ref{step8})
and 
\begin{equation}
\Phi_{q,l}^{diss}\left(\rho_{m}\right)
=\sum_{k=0}^{\infty}A_{k}(t_{0}+l\tau_{pr}+q \tau_{fb})\rho_{m} 
A_{k}(t_{0}+l\tau_{pr}+q \tau_{fb})^{\dagger}
\label{disso}
\end{equation}
describes the standard dissipation acting during the feedback
cycle time $ t_{0}+l\tau_{pr}+q \tau_{fb}$.

\section{Study of the dynamics of the autofeedback scheme}

As we have observed above, the triggering of the feedback cycle 
only when the atomic counters have counted exactly one probe and one
feedback atom makes the time evolution random. In fact, the 
feedback cycle map (\ref{stepind}) we have determined in the preceding 
section is a random map, that is, it depends upon the two
discrete random variables $q$, $l$.
It is evident that if we want to study the dynamics of the microwave mode 
within $C$, two different strategies are possible to determine the
averaged evolution: i) repeat the experiment many times
up to the same, fixed, elapsed time $t$; ii) repeat the experiment 
many times by fixing the number of feedback cycles instead of the
elapsed time. We shall consider this second possibility, in order to better 
understand the effect of the autofeedback scheme. In fact, fixing the
elapsed time would have meant averaging over 
experimental runs characterized
by {\em different} number of feedback cycles. 
Using Eq.~(\ref{stepind}), we have
that in a single run, the state after $N$ feedback cycles is
\begin{equation}
\rho_{N}= \Phi_{q_{N},l_{N}}\Phi_{q_{N-1},l_{N-1}}\ldots
\Phi_{q_{2},l_{2}}\Phi_{q_{1},l_{1}}\rho(0) \;;
\label{mapran2}
\end{equation}
in the limit of a very large number of experimental runs, one gets the
average cavity mode state
\begin{equation}
\bar{\rho}_{N}= \sum p(l_{1})p(q_{1})\ldots p(l_{N})p(q_{N})
\Phi_{q_{N},l_{N}}\Phi_{q_{N-1},l_{N-1}}\ldots
\Phi_{q_{2},l_{2}}\Phi_{q_{1},l_{1}}\rho(0) \;,
\label{mapran3}
\end{equation}
where the probability distributions $p(l)$ are given by Eq.~(\ref{prol}).
Since $q_{1}$, $l_{1} \ldots q_{N}$, $l_{N}$ are independent
random variables, it is evident that the average state $\bar{\rho}_{N}$
after $N$ feedback cycles can also be written as 
\begin{equation}
\bar{\rho}_{N}= \bar{\Phi} ^{N} \rho(0) \;,
\label{mapran4}
\end{equation}
where
\begin{equation}
\bar{\Phi} = \sum p(l)p(q)\Phi_{q,l}
\label{mapmed}
\end{equation}
is the averaged feedback cycle map operator,
determining all the dynamics of the microwave mode.
The expression of this operator can be determined using (\ref{stepind}),
but it is cumbersome and not particularly interesting. One 
relevant aspect of this averaged feedback cycle operator is that, since 
it involves only the even and odd projections $\rho_{g}$ and $\rho_{e}$,
and the cavity mode state is initially confined within the odd subspace,
it never populates the Fock subspace without a definite parity, i.e.,
$\rho_{n,n+p}=0$, whenever $p$ is odd, at all times.
In other words, it is possible to write $\rho=\rho_{g}+\rho_{e}$
at any time.

Let us finally discuss the optimal values of the various 
experimental parameters involved. It is evident that the protection
capabilities of the proposed scheme essentially depend upon the
ratio between the mean feedback cycle time $\bar{t}_{cyc}$ and the 
Schr\"odinger cat decoherence time $t_{dec}= (2 \gamma |\alpha |^{2})^{-1}$.
For smaller and smaller values of this ratio, one gets a longer
and longer protection of the initially generated cat state. This average
cycle time $\bar{t}_{cyc}$ is determined by the spatial dimensions of the
apparatus (which cannot be too miniaturized since we are using 
microwaves) and by the probe and feedback atom velocities, which have to be
therefore as large as possible. However, the probe atom velocity is 
fixed by the $\pi/2$ phase shift condition 
which is needed to have a cat state with a definite parity, 
\begin{equation}
	\frac{\Omega ^{2}t_{int}}{\delta}=
	\frac{\Omega ^{2}L_{C}}{\delta v_{pr}}=\frac{\pi}{2} \;, 
	\label{pifas}
\end{equation}
where $L_{C}$ is the effective transverse length of the $C$ cavity 
mode. In the actual experimental situation $\Omega /2\pi = 24$ kHz, 
$L_{C}=0.75$ cm and the smallest possible value of the detuning, 
compatible with the non-resonant interaction,
is $\delta/2\pi \simeq 70$ kHz, so that
we get $v_{pr}\simeq 250$ m/s. There is no similar
constraint for the feedback atom which can be taken therefore as fast
as possible; we choose $v_{fb} \simeq 500$ m/s, since the Rydberg 
atoms used are thermal Rb atoms and this velocity corresponds to the 
fastest usable part of the Maxwellian distribution. Once we have 
chosen the two atom velocities, one has to check that these values 
are compatible with the $\pi$ pulse condition of Eq.~(\ref{pipul})
for both probe and feedback atom in $C'$ and also with the conditions 
for adiabatic transfer for the feedback atom in $C$. In fact it is 
possible to use the Stark tuning mechanism to determine the 
interaction times in $C'$ satisfying the $\pi$ pulse condition 
only if the cavity crossing time $t_{cr}$ in $C'$ is 
{\em larger} than $\pi/(2 \Omega ')$ [see Eq.~(\ref{pipul})]. Since 
the cavities $C$ and $C'$ are resonant with the two adjacent 
transitions $g \rightarrow e$ and $i \rightarrow g$, they can be 
assumed to be of similar design, so that $\Omega ' \simeq \Omega
=2 \pi \times 24$ kHz and $L_{C'}\simeq L_{C} = 0.75$ cm 
and this implies
$t_{cr}^{pr}\simeq 30$ $\mu$sec, $t_{cr}^{fb}\simeq 15$ $\mu$sec
which are in fact larger than $\pi/(2 \Omega ') \simeq 10$ $\mu$sec.
The condition for the adiabatic passage of the feedback atom in $C$ 
is instead that the feedback atom crossing time in $C$
$t_{cr}^{fb}\simeq 15$ $\mu$sec has to be larger than $\Omega ^{-1}
\simeq 7$ $\mu$sec and therefore this condition is verified too.

The probe and feedback atom velocities determine the overall time of 
flight $t_{0}$ of Eq.~(\ref{step7}); in fact a reasonable estimate of 
the apparatus length from $C$ to $C'$ is $10$ cm and therefore one has
$t_{0} \simeq 600$ $\mu$sec. However, the duration time of a feedback cycle
is determined not only by $t_{0}$, but also by the random waiting times 
$l \tau_{pr}$ and $q\tau_{fb}$ due to the atomic counters 
and also by the non-unit efficiency of the Rydberg state preparation $p_r$
which we can assume to be $p_r \simeq 0.9$.
In fact the probe and feedback atoms are prepared in the correct circular 
Rydberg state with a probability $p_{r}^{2}$ and therefore the photon is 
effectively released into the cavity $C$ only after a random number of
cycles $m$, with probability $p_{A}(m)=p_{r}^{2} (1-p_{r}^{2})^{m-1}$, 
$m=1,2,\ldots$.
As a consequence, the effective mean duration time of a feedback cycle,
that is, the mean time between two successful photon transfers in $C$, is given
by the mean loop time multiplied by the mean number of ``attempts'', 
$\langle m \rangle =1/p_{r}^2$, 
\begin{equation}
	\bar{t}_{cyc}=\langle m \rangle 
	\left(t_{0}+\langle l\rangle \tau_{pr}+
	\langle q\rangle \tau_{fb}\right)
	=\frac{1}{p_{r}^{2}}\left[
	t_{0}+\frac{1-p_{1}}{p_{1}}\left(
	\tau_{pr}+\tau_{fb}\right)\right] \;.
	\label{meanci}
\end{equation}

The sampling time of the probe and feedback atomic counters
$\tau_{pr}$ and $\tau_{fb}$
corresponds to a probed section of the beam of the order of few
millimeters and therefore we can assume $\tau_{pr}
\simeq \tau_{fb} \simeq 15$ $\mu$sec, so that we have $\bar{t}_{cyc}
\simeq 800
$ $\mu$sec. This mean duration time has to be smaller than 
the decoherence time of the Schr\"odinger cat state initially generated,
otherwise the correction of the autofeedback scheme would be too late 
to get a significant protection. However, from the above discussion 
it is evident that the experimental conditions put many constraints on 
the possible parameter values and that this value for $\bar{t}_{cyc}$ 
cannot be significantly decreased. Therefore the only way to achieve 
a significant Schr\"odinger cat state preservation is to increase the 
decoherence time, i.e., increase the relaxation time 
$t_{rel}=\gamma^{-1}$ of the cavity $C$ or decrease the cat state 
initial mean photon number $|\alpha |^{2}$. In fact we can say that a 
cat state $N_{\pm}\left(|\alpha\rangle \pm |-\alpha\rangle \right)$
is protected by the present autofeedback scheme as long as
\begin{equation}
	|\alpha|^{2} < \frac{t_{rel}}{2\bar{t}_{cyc}}\; .
	\label{alcond}
\end{equation}
Alternatively, if we consider a given mesoscopic value for 
$|\alpha|^{2}$, as for example $|\alpha|^{2} = 3.3$ as in 
Ref.~\cite{prlha}, one begins to increase the ``lifetime'' of the 
generated cat state as long as $t_{rel} > 5$ ms.

Relaxation times of this order of magnitude will be hopefully obtained
in the near future and for this reason we have plotted in Fig.~2 the 
Wigner functions and the density matrices describing the averaged 
time evolution in the presence of the autofeedback scheme for an 
initial odd cat state with $\alpha =\sqrt{3.3}$ and for a cavity 
relaxation time $t_{rel}=10$ ms. The cavity $C'$ is assumed to be
equal to $C$ and the values of all the other parameters are the same 
as discussed above, so that $\bar{t}_{cyc}/t_{dec} \simeq 0.53$. 
Fig.~2(a) shows the Wigner function and the density matrix elements
of the initial odd cat state; Fig.~2(b) refers instead to the state of 
the cavity mode after $13$ feedback cycles, corresponding to a mean 
elapsed time $\bar{t} \simeq t_{rel} \simeq 6.6t_{dec}$ and Fig.~2(c) 
refers to the state of 
the cavity mode after $25$ feedback cycles, corresponding to a mean 
elapsed time $\bar{t} \simeq 2t_{rel} \simeq 13t_{dec}$. This figures 
show the impressive preservation of all the main aspects
of the initial odd cat state up to 13 decoherence times. To better 
appreciate the performance of the proposed scheme we show in Fig.~3
the corresponding time evolution of the same initial odd cat state 
in the absence 
of the autofeedback scheme. Fig.~3(a) shows again the initial Wigner 
function and density matrix, Fig.~3(b) refers to the cavity field 
state after one relaxation time $t_{rel}$ and Fig.~3(c) describes the 
cavity field state after two relaxation times (in absence of feedback 
time evolution is no more random and therefore these are actual elapsed
times). In this case, after one relaxation time,
the cat state has already turned into a 
statistical mixture of two coherent states, with no quantum aspect 
left, and it approaches the vacuum state after two relaxation times 
[Fig.~3(c)]. 

Another important aspect of the feedback-induced dynamics shown by 
Fig.~2 is the ``distortion''of the cat state which becomes more and 
more ``rounded'' as time passes. This is due to the slow 
unconventional phase diffusion associated to this feedback scheme and 
which has been discussed in detail in Ref.~\cite{pran}. In fact 
the present autofeedback scheme is an improvement of the original 
scheme of Ref.~\cite{pran}, and the main physical aspects are 
essentially the same: the fed back photon has no phase relationship 
with the photons already present in $C$ and this leads to the 
above mentioned phase diffusion. This phase diffusion turns out to be very 
slow; in fact the present model is essentially a stroboscopic version 
of the continuous photodetection feedback scheme studied in 
\cite{pran}, which is characterized in the semiclassical limit by a 
diffusion term 
\begin{equation}
-\frac{\gamma}{2}\left[\sqrt{n},\left[\sqrt{n},\rho \right]\right]
 \leftrightarrow \frac{\gamma}{8 \bar{n}}\frac{\partial 
^2}{\partial \theta ^2} W(r,\theta ) \;,
\label{derise2}
\end{equation}
for the Wigner function in polar coordinates $W(r,\theta )$ 
($\bar{n}$ is the mean photon 
number) and which is analogous to the phase diffusion of a laser well 
above threshold. It is possible to see (see also Ref.~\cite{pran})
that the asymptotic state of 
the cavity mode is the rotationally invariant mixture of the vacuum 
and the one photon state $\rho_{st}=P_{0}|0\rangle \langle 0|+
P_{1}|1\rangle \langle 1| $, which is however reached after many 
relaxation times.

\section{Conclusions}

In this paper we have proposed a method to 
significantly increase the ``lifetime'' of a Schr\"odinger cat state 
of a microwave cavity mode. The scheme uses ``probe'' and  ``feedback'' 
atoms and a second high-Q microwave cavity to transfer quantum information
between these two atoms without need for a detection stage. This scheme
avoids some of the pitfalls of previously published ones. In particular, its
efficiency does not rely on a perfect Rydberg atom detection. Even though 
it relies on an
efficient preparation of a single atom, this is not critical
since standard laser
techniques can be used to fulfill this requirement. We have shown that 
the method is quite efficient, with realistic orders of magnitude for the 
experimental parameters.

We have focused on the case of a Schr\"odinger cat state which, 
thanks to its well characterized quantum features,  
plays the role of the typical quantum state; 
however, as it can be easily expected, most of the techniques 
presented here could be applied to the case of a generic quantum 
state of a cavity mode (see also Ref.\cite{pran}).

This decoherence control scheme is less general than quantum 
error correction methods because it exploits from the beginning the 
specific aspects of the physical mechanism inducing decoherence. 
However there are similarities between the present 
autofeedback scheme and quantum error correction codes. The 
second cavity $C'$ detects the error syndrome 
during its interaction with the probe atom and sends the necessary 
correction to $C$ via the feedback atom.

After the first experimental evidences of decoherence mechanisms, 
decoherence control
is bound to be a rapidly expanding field in quantum physics. 
First, it is important
as an illustration of a very fundamental relaxation process. 
It would be extremely interesting to tailor decoherence,
as spontaneous emission in the past. This
should lead to a deeper insight into relaxation theory and into 
the border between the microscopic and the macroscopic world.
Decoherence control is also important for quantum information 
processing schemes, since decoherence
is the main problem to manipulate large quantum systems. 
An experimental realization of this feedback scheme,
which is quite realistic, would be an important step in this direction.

\section{Acknowledgments}
This work has been partially supported by INFM (through 
the 1997 Advanced Research Project ``CAT''), by the
European Union in the framework of the TMR Network ``Microlasers
and Cavity QED'' and by MURST under the ``Cofinanziamento 1997''.

\begin{figure}
\centerline{\epsfig{figure=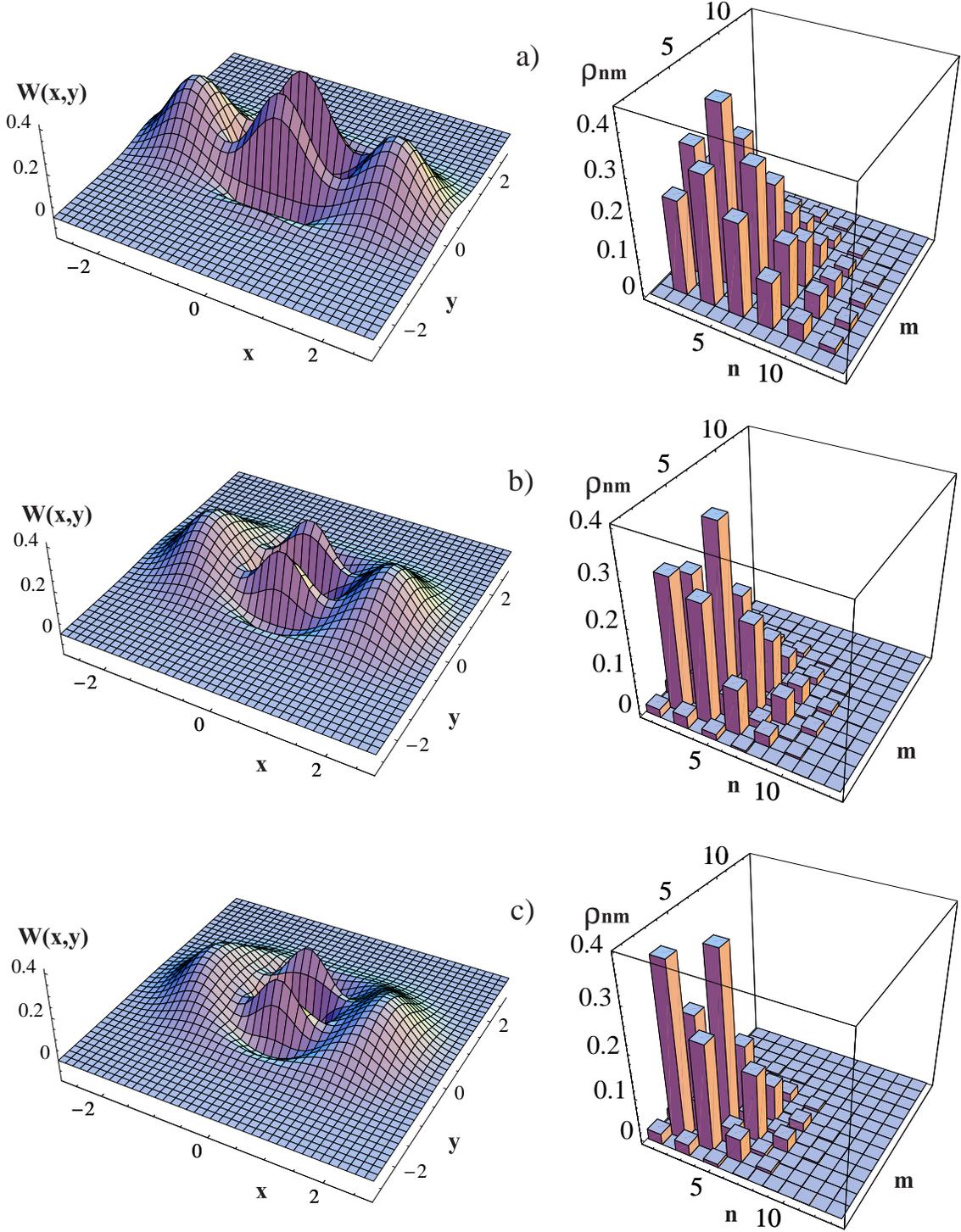,width=15cm}}
\caption{(a) Wigner function and density matrix in the photon number
basis, $\rho_{n,m}=\langle n | \rho | m \rangle$, of the initial odd cat 
state, $|\psi \rangle = N_{-}(|\alpha \rangle - |-\alpha \rangle )$,
$|\alpha |^{2}=3.3$ (b) Wigner function and density matrix $\rho_{n,m}$
of the same cat state after $13$ feedback cycles, corresponding to a mean
elapsed time $\bar{t} \simeq 1/\gamma $ ($\bar{t} \simeq 6.6 t_{dec}$);
(c) Wigner function and density matrix $\rho_{n,m}$ of the same state 
after $25$ feedback cycles  corresponding to a mean elapsed time
$\bar{t} \simeq 2/\gamma $ ($\bar{t} \simeq 13 t_{dec}$). 
All the parameter values are given in the text (see section V).}
\label{strobo1}
\end{figure}

\begin{figure}
\centerline{\epsfig{figure=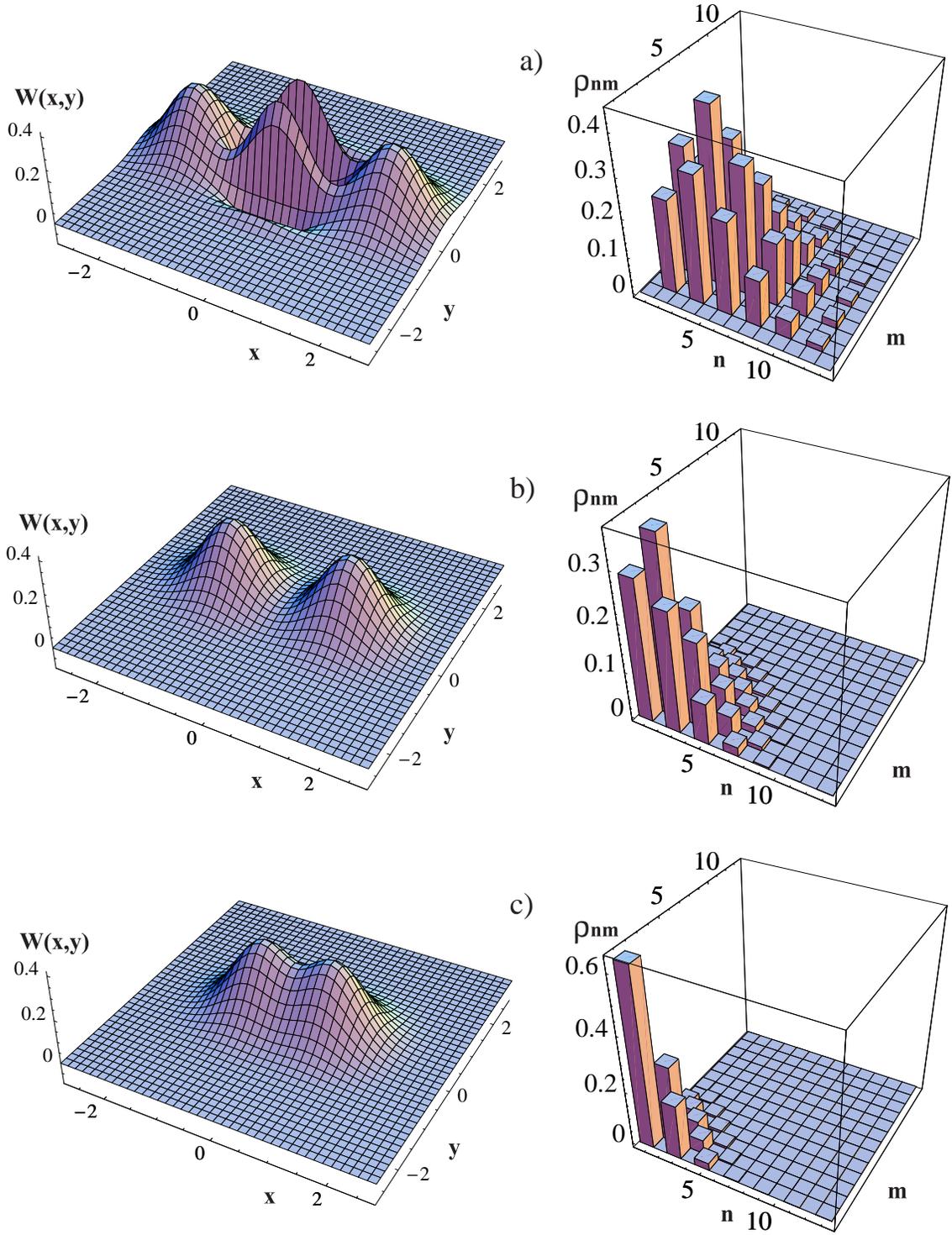,width=15cm}}
\caption{Time evolution of the same initial state of Fig.~2 in 
absence of feedback. (a) Wigner function and density matrix 
in the photon number basis, $\rho_{n,m}=\langle n | \rho | m \rangle$, 
of the initial odd cat state of Fig.~2; (b) Wigner function and density
matrix $\rho_{n,m}$ of the same cat state after one relaxation time
$t=1/\gamma$; (c) Wigner function and density matrix $\rho_{n,m}$ after
two relaxation times  $t=2/\gamma$. The comparison with Fig.~2 is striking:
in absence of feedback the Wigner function becomes quickly positive definite, 
while in the presence of feedback the quantum aspects of the state remain
well visible for many decoherence times.}
\label{strobo2}
\end{figure}

\end{document}